**Original Paper**

**Sharing a common origin between the rotational and linear Doppler effects**

*Liang Fang[1], Miles J. Padgett[2], and Jian Wang[1]\**

*Corresponding Author: E-mail: jwang@hust.edu.cn

[1]Wuhan National Laboratory for Optoelectronics, School of Optical and Electronic Information, Huazhong University of Science and Technology, Wuhan 430074, Hubei, China.
[2] School of Physics and Astronomy, University of Glasgow, Glasgow G12 8QQ, Scotland, UK

**Abstract:** The well-known linear Doppler effect arises from the linear motion between source and observer, while the less well-known rotational Doppler effect arises from the rotational motion. Here, we present both theories and experiments illustrating the relationship between the rotational and linear Doppler effects. A spiral phaseplate is used to generate a light beam carrying orbital angular momentum and the frequency shift is measured arising from its rotational and/or linear motion. By considering either the motion-induced time-evolving phase or the momentum and energy conservation in light-matter interactions, we derive the rotational Doppler shift, linear Doppler shift, and overall Doppler shift associated with rotational and linear motions. We demonstrate the relationship between rotational and linear Doppler shifts, either of which can be derived from the other effect, thereby illustrating their shared origin. Moreover, the close relationship between rotational and linear Doppler effects is also deduced for a more general moving rough surface.

**1. Introduction**

The Doppler effect is the change in frequency of a wave that arises from the relative motion between a source and an observer. The Austrian physicist Doppler first proposed this effect in 1842 [1] and Ballot tested the hypothesis for sound waves in 1845 [2] and Fizeau discovered independently the same phenomenon with electromagnetic waves in 1848 [3]. The Doppler shift arising from either sound waves or electromagnetic waves back-scattered from a moving object is used in sonar and radar systems for velocity measurements. The classical Doppler effect, arising from the linear motion or linear momentum, is also called the linear Doppler effect and has been used extensively in various fields as diverse as laser interferometers, laser remote sensing, laser speckle velocimetry, and astronomy [4-8].

Less well known than this linear Doppler effect is the rotational Doppler effect, which arises from rotational motion and optical angular momentum [9-15]. A simple example of this phenomenon is obtained by placing a watch at the center of a rotating turntable and, as viewed from above, the hands of the watch rotate more quickly than normal [16]. This effect applies not just to watch hands but to all rotating vectors, and the rotating electric-field vector of circularly polarized light, that carries spin angular momentum, is no exception. The electric-field vector rotates at the light frequency, while an additional rotation of the light around its propagation axis can speed up or slow down the rotation of the electric-field vector, leading to a frequency shift linked to the rotational velocity. Note, this rotational Doppler effect can be observed by looking at a rotating body in a direction parallel to its rotation axis, the direction in which the linear shift is zero. This rotational effect is distinct from the linear Doppler effect observed from viewing the edges of a rotating extended body (e.g. galaxy) in a direction perpendicular to its rotation axis.

More generally, the angular momentum of light can be divided into spin angular momentum (SAM) and orbital angular momentum (OAM). The SAM arises from the circular polarization of light and gives two possible values equivalent to $\pm\hbar$ per photon (where $\hbar$ is the Plank's constant $h$ divided by $2\pi$), as anticipated by Poynting in 1909 [17]. By contrast, the OAM arises from helical phase fronts described by $\exp(i\ell\Omega)$ (where $\ell$ is the topological charge and $\theta$ is the azimuthal angle). Unlike the SAM that is restricted to two orthogonal states, the OAM is restricted only by the aperture of the optical system and has a value of $\ell\hbar$ per photon, as recognized by Allen in 1992 [18]. OAM has given rise to many developments in particle manipulation, microscopy, imaging, sensing, astronomy and both quantum and classical communications [19-29]. Recently, OAM-carrying light has also been studied in relation to the rotational Doppler effect, allowing detection of spinning objects, measurement of fluid flow vorticity, and effects in nonlinear optics [30-38].

In this paper, we consider whether the rotational Doppler effect and linear Doppler effect share a common origin, unlike for example the relativistic transverse Doppler effect which is related to time dilation. By employing a spiral phaseplate and exploiting OAM-carrying light, we illustrate the relationship between rotational and linear Doppler effects, i.e. show how either frequency shift can be alternatively explained both in terms of the rotational or linear Doppler effect.

## 2. Concept and Theory

### 2.1. Spiral Phaseplate

A natural way of generating light beams containing OAM is by transmission of a plane-wave Gaussian beam through a spiral phaseplate [39]. A spiral phaseplate is simply a transmissive disc with an optical thickness, $H$, that increases with azimuthal angle, $H = \ell \lambda \cdot \theta/(2\pi)$, where $\lambda$ is the wavelength of light. Instead of transmission, as shown in Fig. 1, the same concept can be also applied in reflection where an incident plane wave is reflected from a helical surface with a step height of $H = \ell \lambda/2 \cdot \theta/(2\pi) = \ell\lambda\theta/(4\pi)$, giving a reflected light beam with a helical phase front described by $\exp(i\ell\theta)$, and hence an OAM of $\ell\hbar$ per photon. The factor of two reduction in step height between the two equations when moving from transmission to reflection is to account for the double pass associated with the reflection. As an alternative to a physical implementation, these spiral phaseplates can be implemented using the spatially dependent geometrical phase associated with transmission through liquid crystal [40] or metamaterials [41], or encoded directly onto a spatial light modulator (SLM) [19, 20, 42].

## 2.2 Rotational Doppler Effect

As shown in Fig. 1(a), if the spiral phaseplate is rotated through an angle $\Delta\theta$ around its central axis then the reflected light acquires a phase shift of $\Delta\varphi = \ell \cdot \Delta\theta$. If this rotation is continuous, with rotational velocity $\Omega = d\theta/dt$, then the resulting time-evolving phase is manifest as a frequency shift in the reflected light of $\Delta f = d\varphi/dt/(2\pi) = \ell\Omega/(2\pi)$, a form of the rotational Doppler shift.

This rotational Doppler shift can also be understood from the conservation of the angular momentum and energy in the light-matter interaction [43]. Upon reflection the light has a change in its angular momentum of $\ell\hbar$ per photon. Accordingly, the conservation of angular momentum dictates that the rotating spiral phaseplate experiences an opposite impulse torque of $\ell\hbar$ per photon. The spiral phaseplate rotating at rotational velocity $\Omega$ does work against this torque delivering an energy to the reflected light of $\Delta E = \Omega\ell\hbar$ per photon, leading to a frequency shift of $\Delta f = \Delta E / h = \ell\Omega/(2\pi)$, i.e. the rotational Doppler shift.

## 2.3 Linear Doppler Effect

The linear Doppler shift is also understandable based on similar conservations of linear momentum and energy in the light-matter interaction. A single photon reflected from the surface undergoes a change in its linear momentum of $2k\hbar$, with the mirror being subject to an equal but opposite impulse, where $k = 2\pi/\lambda = 2\pi f/c$ is the wavenumber, $f$ is the frequency of light, and $c$ the light velocity. If the mirror is moving at linear velocity $v$, this impulse corresponds to an energy transfer per photon in the reflected light of $\Delta E = 2kv\hbar$. Equating this energy transfer to the change in optical frequency gives a frequency shift of $\Delta f = \Delta E / h = 2fv/c$, i.e. the linear Doppler shift (Note that in this case the factor of two arises because the light is Doppler shifted once at the surface and then Doppler shifted a second time as it is received from this moving surface back at the source).

**2.4 Linking the Rotational and Linear Doppler Effects**

In the case of the spiral phaseplate, the rotational frequency shift can also be alternatively derived from the linear Doppler effect. As shown in Fig. 1(b), the back reflection from a planar mirror moving with respect to the source with a linear velocity $v$, creates a time-evolving phase of $2kvt$, resulting in a linear Doppler shift given by $\Delta f = d(2kvt)/dt/(2\pi) = 2fv/c$ as above. The rotation of the spiral phaseplate also creates a movement of its helical surface towards or away from the source/observer. For a step height of $\ell\lambda/2$ and a rotational velocity of $\Omega$, shown in Fig. 1(a), any location on the helical surface of the phaseplate has a linear velocity with respect to the source of $v = d[(\ell\lambda/2)\cdot\theta/(2\pi)]/dt = \ell\lambda/(4\pi)\cdot d\theta/dt = \Omega\ell\lambda/(4\pi)$. This linear velocity gives an observed frequency shift of $\Delta f = 2fv/c = 2f\cdot\Omega\ell\lambda/(4\pi)/c$, or simply $\Delta f = \ell\Omega/(2\pi)$, i.e. identical to that derived from the rotational Doppler shift. Consequently, here is an example where a frequency shift can be equivalently explained either in terms of the rotational Doppler effect or linear Doppler effect, illustrating, at least in this case, their common origin.

Remarkably, it is well known that, the rotational Doppler shift $\Delta f = \ell\Omega/(2\pi)$ is frequency independent, while the linear Doppler shift $\Delta f = 2fv/c$ is dependent on the frequency of light, giving a false impression of their conflict in the shared common orgin. Actually, when deducing the rotational Doppler shift based on the linear Doppler effect, the equivalent linear velocity $v = \Omega\ell\lambda/(4\pi)$ of any location on the helical surface of the rotating spiral phaseplate ($\Omega$) is also frequency (wavelength) dependent. As a result, the frequency dependence in the linear Doppler shift is removed in the deduced rotational Doppler shift.

The rotational Doppler effect can be deduced from the linear Doppler effect in the case of the spiral phaseplate and vice versa. The linear motion ($v$) of the spiral phaseplate, causing a phase change, is also equivalent to its rotational motion ($\Omega = 4\pi v/(\ell\lambda)$) producing the same amout of the phase variation. Hence, the linear Doppler shift can also be alternatively derived from the rotational Doppler effect.

**2.5 Overall Doppler Shift**

When the spiral phaseplate is subject to simultaneous rotational motion and linear motion, as shown in Fig. 1(c), a total frequency shift can also be deduced. The incident plane wave is transformed into an OAM-carrying light beam by the spiral phaseplate, meaning that the reflected light undergoes a change in its angular momentum of $\ell\hbar$ per photon and additionally the reflection results in a change in its linear momentum of $2k\hbar$. Hence, the moving spiral phaseplate experiences both a impulse torque of $\ell\hbar$ and radiation pressure impulse of $2k\hbar$ per photon. Therefore moving the spiral phaseplate with rotational velocity $\Omega$ against this impulse torque and with linear velocity $v$ against this radiation pressure, produces a total energy transfer to the reflected light of $\Delta E = (\ell\Omega + 2kv)\hbar$ per photon, resulting in a frequency shift of $\Delta f = \Delta E/h = \ell\Omega/(2\pi) + 2fv/c$. The overall Doppler shift is therefore the sum of both the rotational Doppler shift ($\ell\Omega/(2\pi)$) and the linear Doppler shift ($2fv/c$). By linking the rotational and linear Doppler effects, we can simply obtain the same results of equivalent rotational or linear Doppler shift irrespective of the approach taken to their individual calculation (see Supporting Information).

It is clear that the rotational Doppler effect and linear Doppler effect have close relationship. The overall Doppler shift and the relationship between rotational Doppler shift and linear Doppler shift can be also deduced by considering the moving spiral phaseplate along the angular direction as a triangular wedge in polar coordinates (see Supporting Information).

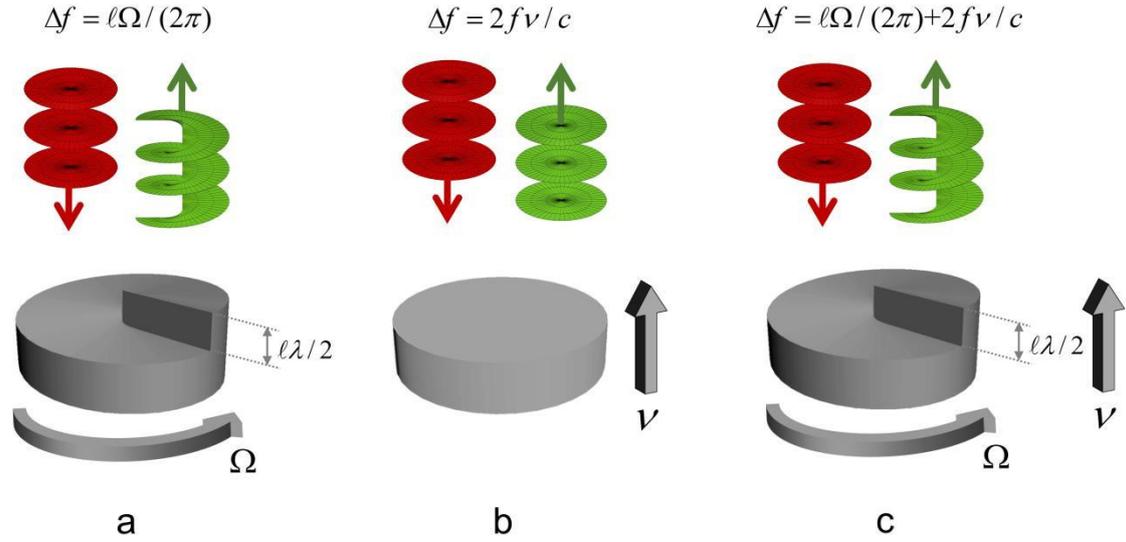

**Figure1 Illustration of rotational Doppler effect, linear Doppler effect and overall Doppler effect.** (**a**) A spiral phaseplate rotating around the optical axis gives rise to the rotational Doppler shift. (**b**) A mirror translating along the optical axis gives rise to the linear Doppler shift. (**c**) A moving spiral phaseplate with simultaneous rotational motion and linear motion gives rise to the overall Doppler shift (rotational Doppler shift and linear Doppler shift). Both of these shifts can be understood in terms of the linear Doppler shift albeit the rotational case is a specific example of a general rotational Doppler shift.

## 3. Experimental Realization

### 3.1 Experimental Configuration

To verify the link between the rotational and linear Doppler effects, we setup an experimental configuration, as shown in Fig. 2(a). The configuration can be regarded as a modified Mach-Zehnder interferometer with the scanning mirror replaced with the moving spiral phaseplate. An SLM loaded with a time-varying spiral phase mask is employed to emulate the moving spiral phaseplate (see Supporting Information). A He-Ne laser emitting at 632.8 nm is divided into two optical paths by a non-polarizing beam splitter (BS). One path is used as a reference Gaussian light beam with its power and beam size adjusted by a neutral density filter (NDF) and a lens, respectively. The other path is reflected from the time-varying spiral phase mask through a mirror and a second BS. A polarizer (Pol.) and a half-wave plate (HWP) after the laser are used to set the polarization state of the light to match the required polarization of the polarization-dependent SLM. The OAM-carrying light beam reflected from the spiral phase mask encoded on the SLM is reflected by the second BS and then combined with the reference Gaussian light beam using a further BS together to produce the characteristic helical interference fringes [44]. Another lens followed by a camera enables an image of the helical interference fringe pattern to be recorded, as shown in Fig. 2(b), from which the sign and magnitude of the OAM is confirmed and, from the rotation of the fringes, the Doppler frequency shift is deduced (see Supporting Information).

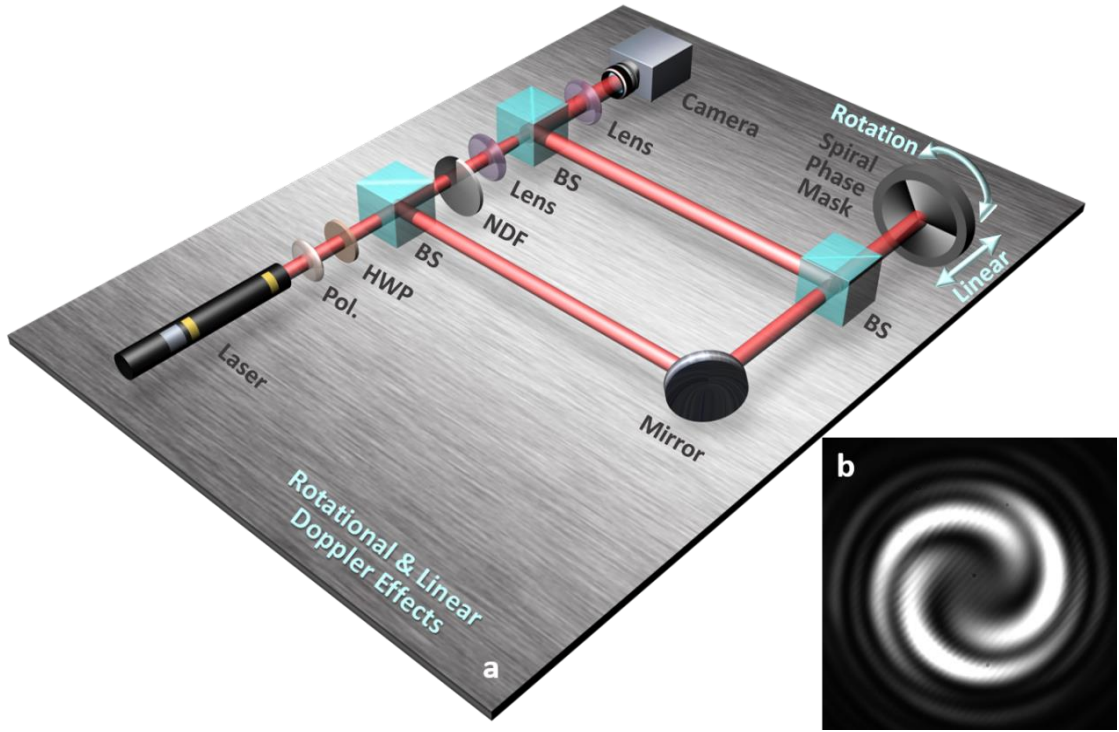

**Figure 2 Experimental configuration and helical interference fringes.** (**a**) Experimental configuration of a modified Mach-Zehnder interferometer for the measurement of rotational and linear Doppler effects. The rotational motion and linear motion of the spiral phaseplate are emulated by employing a spatial light modulator (SLM) loaded with a time-varying spiral phase mask. Pol.: polarizer; HWP: half-wave plate; BS: beam splitter; NDF: neutral density filter. (**b**) Measured helical interference fringes (interference between an OAM-carrying light beam and a reference Gaussian light beam). The twist number and twist direction of the helical interference fringes are determined by the magnitude and sign of the OAM, and their rotation is a manifestation of the frequency shift.

### 3.2 Results and Discussions

To highlight the equivalence of rotational Doppler shift and linear Doppler shift, we measure the frequency shift arising from separate rotational motion and linear motion. To confirm the derived relationship of $v = \Omega \ell \lambda /(4\pi)$ or $\Omega = 4\pi v /(\ell \lambda)$, we choose two initial cases where case 1 is the rotational only motion ($\Omega = 500$ rad/s, $v = 0$) and case 2 is the linear only motion ($\Omega = 0$, $v = 76$ μm/s).

As a representative OAM-carrying light beam we report the results for $\ell = 3$ and in both cases obtained by measuring the time-varying intensity at an off-axis position in the helical interference fringes. The recorded time-varying intensity and its Fourier-transformed frequency spectra for two cases are shown in Fig. 3. One can see clearly that the measured rotational Doppler shift of $\Delta f \approx 241$ Hz is almost identical to the measured linear Doppler shift of $\Delta f \approx 240$ Hz. Thereby, unsurprisingly, the same frequency shift can be produced by either rotational or linear Doppler effect.

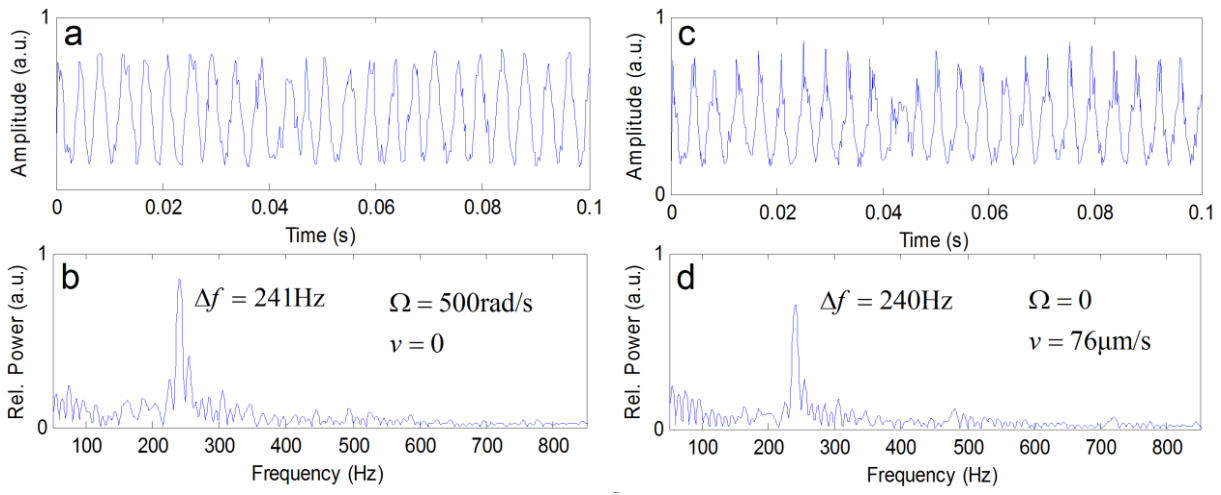

**Figure 3 Time-varying intensity and Doppler frequency shift.** (**a**), (**c**) Measured time-varying intensities at an off-axis position in the helical interference fringes. (**b**), (**d**) Fourier-transformed frequency spectra. (**a**), (**b**) Case 1: rotational only motion ($\Omega = 500$ rad/s, $v = 0$) with measured frequency shift of $\Delta f \approx 241$ Hz. (**c**), (**d**) Case 2: linear only motion ($\Omega = 0$, $v = 76$ μm/s) with measured frequency shift of $\Delta f \approx 240$ Hz. The OAM value is $\ell = 3$.

To further compare the rotational and linear Doppler shifts and show their close relationship, we employ different OAM values of $\ell = 1, 3, 5$ and different rotational and linear velocities. Shown in Fig. 4 is the measured frequency shift as a function of the rotational velocity or linear velocity. It is clearly shown that the $\Delta f - v$ curve is independent on OAM values while, as expected, different OAM values give different slopes of $\Delta f - \Omega$ curves. The larger the OAM value, the larger the slope of the $\Delta f - \Omega$ curve. Shown in the inset of Fig. 4 are $v$-$\Omega$ curves indicating the relationship between linear velocity and rotational velocity ($v = \Omega \ell \lambda / (4\pi)$) required to give the same Doppler frequency shift. Consequently, the same frequency shift can be generated from either the rotational Doppler effect or the linear Doppler effect.

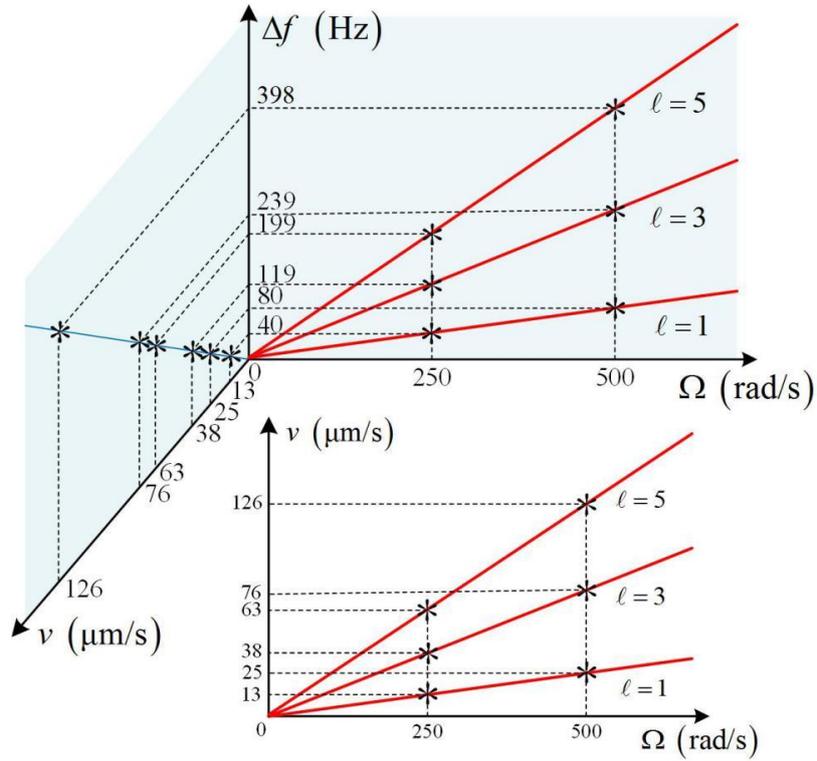

**Figure 4 Rotational and linear Doppler shifts.** Measured frequency shift versus rotational velocity ($\Omega$) or linear velocity ($v$). $\Delta f$-$v$ curve is independent on OAM values. Different OAM values give different slopes of $\Delta f$-$\Omega$ curves. The inset shows $v$-$\Omega$ curves. Markers: experimental results; solid lines: theories ($\Delta f = \ell \Omega / (2\pi)$, $\Delta f = 2 f v / c$, $v = \Omega \ell \lambda / (4\pi)$).

For the overall Doppler shift arising from simultaneous rotational motion and linear motion shown in Figs. 5(a) and 5(d), following the relationship of $v = \Omega \ell \lambda / (4\pi)$ or $\Omega = 4\pi v / (\ell \lambda)$ between the rotational velocity and linear velocity, we also show its equivalent linear Doppler shift with linear only motion shown in Figs. 5(b) and 5(e) and equivalent rotational Doppler shift with rotational only motion shown in Figs. 5(c) and 5(f).

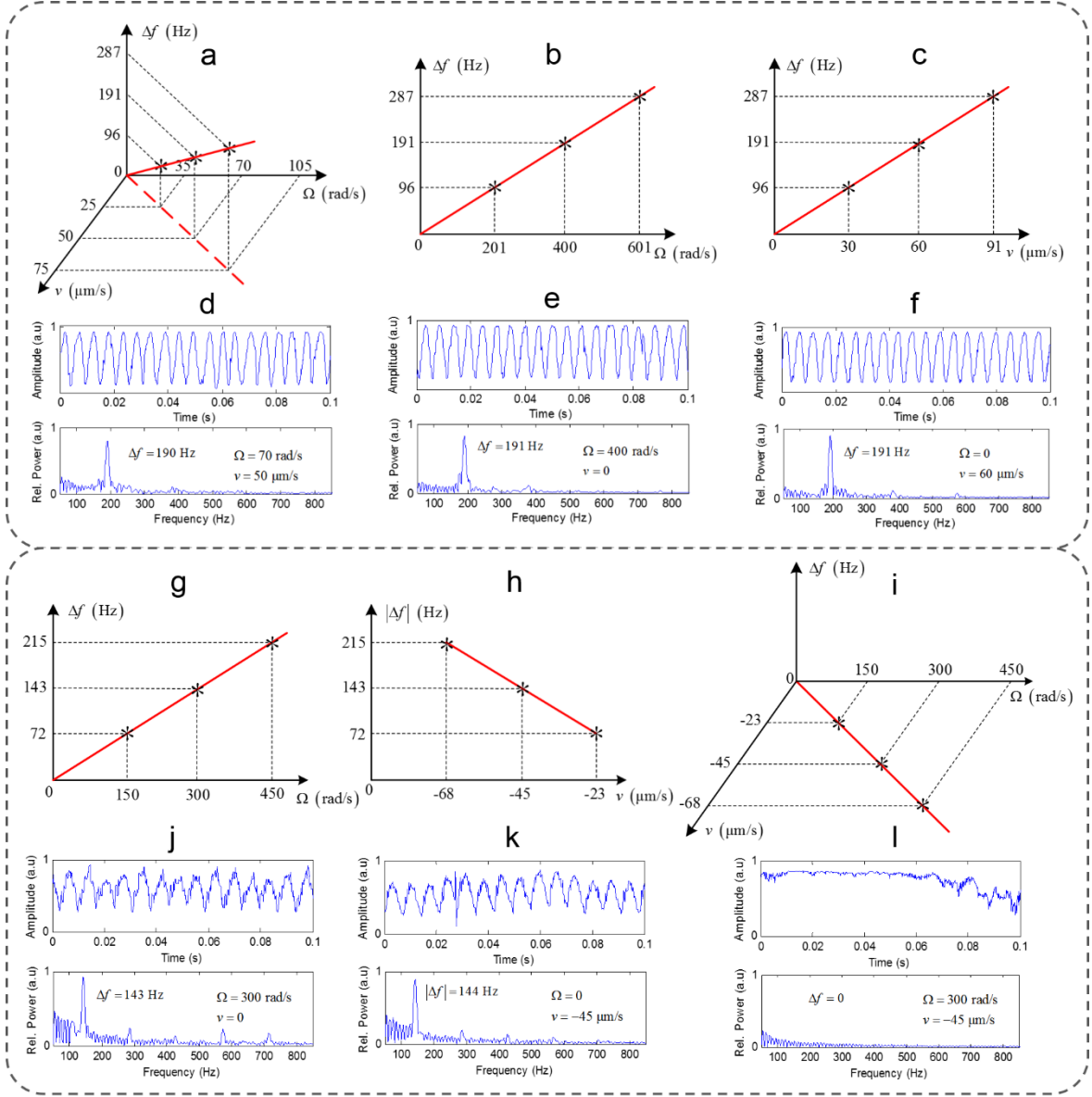

**Figure 5 | (a)-(f) Overall Doppler shift and its equivalent rotational and linear Doppler shift. (g)-(l) Overall Doppler shift close to zero.** (**a**) Measured overall Doppler shift versus rotational velocity ($\Omega$) and linear velocity ($v$) for simultaneous rotational motion and linear motion. (**b**) Measured equivalent rotational Doppler shift versus rotational velocity ($\Omega$) for rotational only motion. (**c**) Measured equivalent linear Doppler shift versus linear velocity ($v$) for linear only motion. (**d**), (**e**), (**f**) Measured time-varying intensities at an off-axis position in the helical interference fringes (upper row) and Fourier-transformed frequency spectra (lower row) corresponding to the middle point in (**a**), (**b**), (**c**). (**g**) Measured rotational Doppler shift versus rotational velocity ($\Omega$) for rotational only motion. (**h**) Measured linear Doppler shift versus linear velocity ($v$) for linear only motion. (**i**) Measured overall Doppler shift close to zero for simultaneous rotational motion and linear motion satisfying the relationship of $v = -\Omega \ell \lambda /(4\pi)$. (**j**), (**k**), (**l**) Measured time-varying intensities at an off-axis position in the helical interference fringes (upper row) and Fourier-transformed frequency spectra (lower row) corresponding to the middle point in (**g**), (**h**), (**i**). The OAM value is $\ell = 3$. Markers: experimental results; solid lines: theories ((**a**), (**i**) $\Delta f = 2 f v / c + \ell \Omega /(2\pi)$, (**b**), (**g**) $\Delta f = \ell \Omega /(2\pi)$, (**c**), (**h**) $\Delta f = 2 f v / c$).

Obviously, the rotational Doppler shift and the linear Doppler shift can also cancel with each other, resulting in an overall Doppler shift close to zero, as shown in Figs. 5(g)-5(l). Although separate rotational/linear only motion is detectable from its corresponding rotational/linear Doppler shift shown in Figs. 5(g), 5(h), 5(j) and 5(k), the overall motion is undetectable when the rotational velocity and linear velocity satisfy the relationship of $v=-\Omega\ell\lambda/(4\pi)$, as shown in Figs. 5(i) and 5(l).

The measured data in the proof-of-concept experiment shown in Figs. 3-5 are in good agreement with the theories. The maximum relative error of the measurement is less than 5%. The obtained results show successful demostration of the close relationship between rotational and linear Doppler effects. High precision measurement is achieved in the experiment when using SLM as an alternative to the physical implementation of a spiral phaseplate (see Supporting Information).

## 4. General Rough Surface

More generally, we consider the equivalent relationship between the rotational and linear Doppler effects for scattered light from a moving rough surface. As illustrated in Fig. 6, the incident light beam is scattered by a moving rough surface with simultaneous rotational motion and linear motion. The scattered light can be decomposed into a sum of different OAM components. A specific inverse spiral phase mask can be used to collect the desired OAM component and convert it into a Gaussian-like light beam with a bright spot at the beam center. The collected scattered light is then sent to a collimator (Col.) followed by a single-mode fiber (SMF) for spatial filtering and detection.

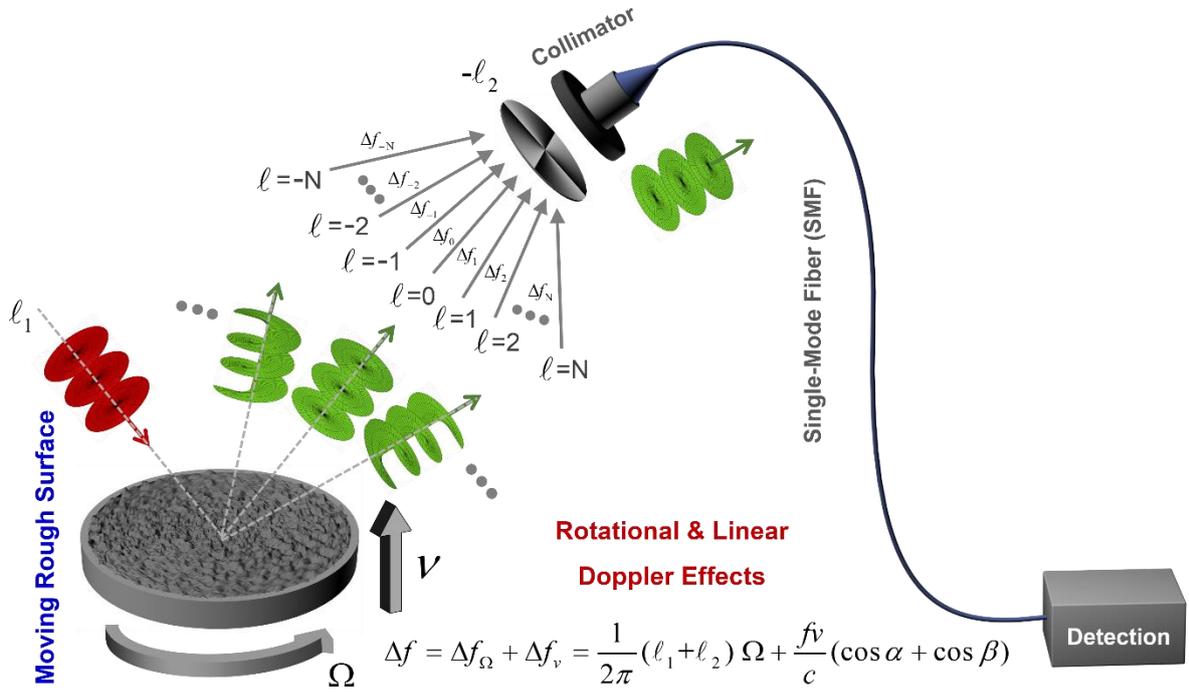

**Figure 6 Illustration of Doppler effects arising from a general moving rough surface.** Rotational and linear Doppler shifts ($\Delta f_\Omega$, $\Delta f_v$) for scattered light from a moving rough surface with rotational motion (rotational velocity $\Omega$) and linear motion (linear velocity $v$). The rough surface can be expanded by a series of helical phase components. The scattered light contains a sum of OAM components. The collected scattered light (one OAM component $\ell_2$) by a specific inverse phase mask ($-\ell_2$) is sent to a collimator (Col.) and a single-mode fiber (SMF) for detection. $\alpha$ and $\beta$ denote the incident light angle and scattered light angle (detection angle) with respect to the normal of the rough surface.

A rough surface can be described as a reflective transfer function $H(r,\theta)$ in polar coordinates with $r$ the radius and $\theta$ the azimuthal angle. It can be expanded by a series of helical phase components $H(r,\theta) = \sum_m B_m(r)\exp(im\theta)$, where $B_m(r)$ is the modulation coefficient of each component, and can be normalized with $\sum_m \left| \int B_m(r)\,dr \right|^2 = 1$. When the rough surface moves with a rotational velocity $\Omega$ and a linear velocity $v$ (the direction of the linear motion is parallel to the normal of the rough surface), the modified transfer function of the moving rough surface can be written by

$$H(r,\theta,v,\Omega) = \sum_m B_m(r)\exp\left[im(\theta+\Omega t)\right]\exp\left[i(\cos\alpha + \cos\beta)kvt\right] \qquad (1)$$

where $k$ is the wavenumber, $\alpha$ and $\beta$ indicate the incident light angle and scattered light angle (detection angle) with respect to the normal of the rough surface, respectively. When facing the rough surface, the sign (+/-) of $\Omega$ and $v$, denotes the clockwise/counterclockwise rotational motion and forward/backward linear motion, respectively.

The electric field of an incident light beam, carrying an OAM of $\ell_1$ at a frequency $f$, is given by

$$E_1(r,\theta,t) = C_{\ell_1} \exp\left[i(2\pi ft + \ell_1\theta)\right] \quad (2)$$

where $C_{\ell_1}$ is related to the transverse electric field distribution of the OAM-carrying light beam and the wavenumber $k$ and propagation distance $z$ related phase term is not given here. The scattered light from the moving rough surface can be expressed as

$$\begin{aligned} E_2(r,\theta,v,\Omega,t) &= E_1(r,-\theta,t) \cdot H(r,\theta,v,\Omega) \\ &= \sum_m B_m C_{\ell_1} \exp\left[i(m-\ell_1)\theta\right] \exp\left\{i\left[2\pi f + m\Omega + (\cos\alpha + \cos\beta)kv\right]t\right\} \end{aligned} \quad (3)$$

It clearly shows that for the scattered light, different OAM ($\ell_2 = m - \ell_1$) components have different overall Doppler frequency shifts, generally written by

$$\Delta f = \Delta f_\Omega + \Delta f_v = \frac{1}{2\pi}m\Omega + \frac{fv}{c}(\cos\alpha + \cos\beta) \quad (4)$$

For separate rotational only motion, the corresponding rotational Doppler shift is expressed as

$$\Delta f_\Omega = \frac{1}{2\pi}m\Omega \quad (5)$$

For separate linear only motion, the corresponding linear Doppler shift is expressed as

$$\Delta f_v = \frac{fv}{c}(\cos\alpha + \cos\beta) \quad (6)$$

For the rotational motion with a rotational velocity of $\Omega$, the resultant rotational Doppler shift can be also understood from the same amount of linear Doppler shift by a linear motion with the corresponding linear velocity $v$ written by

$$v = \frac{\Omega}{2\pi} \frac{m\lambda}{\cos\alpha + \cos\beta} \quad (7)$$

Similarly, for the linear motion with a linear velocity of $v$, the resultant linear Doppler shift can be also understood from the same amount of rotational Doppler shift by a rotational motion with the corresponding rotational velocity $\Omega$ written by

$$\Omega = \frac{2\pi v}{m\lambda}(\cos\alpha + \cos\beta) \quad (8)$$

For the OAM component $\ell_2 = -\ell_1$ ($m=0$) in the scattered light, it is clear from Eqs. (4) and (5) that there is no rotational Doppler shift, which is analogous to the case of no rotational Doppler shift for mirror reflection with an incident OAM beam.

We consider two more special cases of Doppler shift for scattered light from a moving rough surface: 1) the incident light beam is a plane wave while the collected scattered light component carries an OAM ($\ell \neq 0$); 2) the incident light beam carries an OAM ($\ell \neq 0$) while the collected scattered light component is a plane wave. The former is corresponding to the aforementioned spiral phaseplate case, while the latter interchanges the source and observer. In general, the incident angle $\alpha$ and scattered angle $\beta$ may take moderate values. When assuming small incident and scattered angles for both two cases ($\alpha \approx \beta \approx 0$, $\cos\alpha = 1 - 2\sin^2(\alpha/2) \approx 1 - \alpha^2/2 \approx 1$, $\cos\beta = 1 - 2\sin^2(\beta/2) \approx 1 - \beta^2/2 \approx 1$), the overall Doppler shift expressed in Eq. (4) can be simplified as $\Delta f = \ell\Omega/(2\pi) + 2fv/c$. Accordingly, the rotational Doppler shift for separate rotational only motion is $\Delta f_\Omega = \ell\Omega/(2\pi)$, and the linear Doppler shift for separate linear only motion is $\Delta f_v = 2fv/c$. The rotational Doppler shift under a rotational velocity of $\Omega$ is also understandable from the linear Doppler shift under an equivalent linear velocity $v = \Omega\ell\lambda/(4\pi)$. The linear Doppler shift under a linear velocity of $v$ is also understandable from the rotational Doppler shift under an equivalent rotational velocity $\Omega = 4\pi v/(\ell\lambda)$. These show exactly the same results as obtained above for a moving spiral phaseplate or its equivalent physical implementations.

Remarkably, when expanding the general rough surface into a series of helical phase components, each component is equivalent to a spiral phaseplate. According to Eq. (3), for each OAM component in the scattered light (note that the collected scattered light selects one OAM component using a specific inverse spiral phase mask), the moving rough surface is indeed equivalent to a corresponding moving spiral phaseplate or its equivalent physical implementations. The common origin of rotational and linear Doppler effects is understandable from the moving spiral phaseplate, and so is the general moving rough surface. From a more general point of view, both rotational Doppler shift and linear Doppler shift arise from motion-induced time-evolving phase of collected scattered light. Consequently, the same frequency shift, originated from the motion-induced time-evolving phase of collected scattered light, can be equivalently explained by either rotational Doppler effect or linear Doppler effect, implying their common origin even for the general moving rough surface.

## 5. Conclusion

In summary, we use an SLM to implement a spiral phaseplate to generate an OAM-carrying light beam and consider both its rotational and linear motions to link the rotational and linear Doppler effects. By either analyzing motion-induced time-evolving phase or according to the momentum and energy conservation in the light-matter interaction, we show two different ways for how the rotational Doppler shift arises from the rotational motion and for how the linear Doppler shift arises from the linear motion. For the simple example of a rotating and linearly moving spiral phaseplate or other equivalent physical implementations, we show, both theoretically and experimentally, that rotational Doppler effect and linear Doppler effect can be deduced from each other. Both the rotational and linear frequency shifts can be explained by either the rotational Doppler effect or the linear Doppler effect, illustrating their common origin. We also discuss the overall Doppler shift associated with simultaneous rotational motion and linear motion. We further show that the rotational Doppler effect and linear Doppler effect, can also cancel with each other.

For a more general moving rough surface, we expand the rough surface into a series of helical phase components, each being equivalent to a spiral phaseplate. For each OAM component in the scattered light, we show clearly the overall Doppler shift, rotational Doppler shift, linear Doppler shift and close relationship between rotational and linear Doppler effects, illustrating their common origin even for the general moving rough surface.

This study provides a new understanding on the close relationship between rotational and linear Doppler effects. The demonstrations may open up new perspectives to more extensive applications in optical sensing and optical metrology exploiting linear Doppler effect, rotational Doppler effect and overall Doppler effect.

**Supporting Information**

Additional supporting information may be found in the online version of this article at the publisher's website.

**Acknowledgements**  This work was supported by the National Natural Science Foundation of China (NSFC) (11774116, 61761130082, 11574001, 11274131), the Royal Society-Newton Advanced Fellowship, the National Programm for Support of Top-notch Young Professionals, and the National Basic Research Programm of China (973 Program) (2014CB340004).